\documentclass[aps,prb,twocolumn,superscriptaddress,fleqn]{revtex4-1}  % for review and submission
\usepackage[english]{babel}
\usepackage{amsmath}
\usepackage{amssymb}
\usepackage{graphicx} 
\usepackage{siunitx}
\usepackage{ulem}
\usepackage{booktabs}

\usepackage[usenames, dvipsnames]{color}

\begin{document}

\title{Nonlinear plasma wavelength scalings in a laser wakefield accelerator}
\author{H. Ding}
\author{A. D\"opp}
\affiliation{Ludwig-Maximilians-Universitat M\"unchen, Am Coulombwall 1, 85748 Garching, Germany}
\author{M. Gilljohann}
\affiliation{Ludwig-Maximilians-Universitat M\"unchen, Am Coulombwall 1, 85748 Garching, Germany}
\author{J. G\"otzfried}
\affiliation{Ludwig-Maximilians-Universitat M\"unchen, Am Coulombwall 1, 85748 Garching, Germany}
\author{S. Schindler}
\affiliation{Ludwig-Maximilians-Universitat M\"unchen, Am Coulombwall 1, 85748 Garching, Germany}
\author{L. Wildgruber}
\affiliation{Ludwig-Maximilians-Universitat M\"unchen, Am Coulombwall 1, 85748 Garching, Germany}
\author{G. Cheung}
\affiliation{John Adams Institute \& Department of Physics, Clarendon Laboratory, University of Oxford, Parks Road,
Oxford OX1 3PU, United Kingdom}
\author{S. M. Hooker}
\affiliation{John Adams Institute \& Department of Physics, Clarendon Laboratory, University of Oxford, Parks Road,
Oxford OX1 3PU, United Kingdom}
\author{S. Karsch}
\affiliation{Ludwig-Maximilians-Universitat M\"unchen, Am Coulombwall 1, 85748 Garching, Germany}

\begin{abstract}
Laser wakefield acceleration relies on the excitation of a plasma wave due to the ponderomotive force of an intense laser pulse. However, plasma wave trains in the wake of the laser have scarcely been studied directly in experiments. Here we use few-cycle shadowgraphy in conjunction with interferometry to quantify plasma waves excited by the laser within the density range of GeV-scale accelerators, i.e.~a few \num{e18} \si{\per \cubic \centi \metre}. While analytical models suggest a clear dependency between the non-linear plasma wavelength and the peak potential $a_0$, our study shows that the analytical models are only accurate for driver strength $a_0\lesssim 1$. Experimental data and systematic particle-in-cell simulations reveal that nonlinear lengthening of plasma wave train depends not solely on the laser peak intensity but also on the waist of the focal spot.
\end{abstract}

\maketitle
Laser wakefield accelerators (LWFAs) have seen tremendous development since their inception in the late 1970s\cite{Tajima:1979un}. Current LWFAs have reached up to multi-GeV beam energies\cite{Wang:2013el,Kim:2017il, Gonsalves:2019ht}, controlled injection schemes have drastically increased their stability and tunability \cite{Faure:2006vy, McGuffey:2010wy, Buck:2013gs, Thaury:2015dq, Wenz:2019gc}, and several types of LWFA-based compact X-ray sources have achieved competitive peak brightness compared with RF-technology-based infrastructures\cite{Corde:2013bja, Khrennikov:2015gxa, Dopp:2017dza}. Furthermore, these sources have demonstrated their application potential in X-ray imaging\cite{Wenz:2015if, Dopp:2018ub}, high energy density physics \cite{Mahieu:2018bx, Wood:2018gs}, and tumor treatment\cite{Oppelt:2015DoseResponse}.

While the field of wakefield acceleration is developing at a fast pace, some basic questions are still not fully answered. In particular, though theories on the formation of plasma wave trains have been studied extensively for the one-dimensional case, their predicting power in a real experiment is often limited due to higher dimensional effects. Studies addressing 3D plasma wave formation date back decades and remain mostly qualitative \cite{Bulanov:1997ut} or phenomenological\cite{Lu:2007eb}. \citet{Lu:2006ja, Lu:2006ii} established a quantitative model to correlate plasma bubble size with the laser peak intensity based on a force balance argument, which, however, is only valid in the bubble regime and does not discuss trailing periods of plasma oscillations.

Due to the restrictions of analytical models, interpretation of experimental results relies heavily on numerical simulations. Recent development of fast particle-in-cell codes such as CALDER-CIRC\cite{Lifschitz20091803} and FBPIC\cite{Lehe:2016dn} allows to perform quasi-3D simulations in a short period of time, thus enabling systematic parameter scans.

Furthermore, new plasma diagnostics such as few-cycle shadowgraphy uniquely combine femtosecond resolution with picosecond observation windows \cite{Downer:2018ck, Gilljohann:2019kc}. As established models predict a clear relation between the plasma wavelength and the laser peak potential, this method potentially provides a novel non-invasive diagnostic for the laser evolution. Pioneering work of \citet{Savert:2015jr} has demonstrated the lengthening of the plasma bubble\cite{Pukhov:2002ts} and provided important information about the electron injection process.

This manuscript is structured as follows: First, we will revisit established models which predict changes on the plasma wavelength in dependence of the laser intensity. This is followed by a presentation of experimental data from few-cycle shadowgraphy and interferometry performed at the ATLAS laser in Garching, Germany. Afterwards we present the result of systematic particle-in-cell studies, which give deeper insight into the scalings of the plasma wavelength with laser intensity and waist.

\section{Theory}

A wakefield potential $\Phi$ generated behind a laser pulse can be described by a 1D perturbative fluid theory as
\begin{equation}
    \left( \frac{\partial^2}{\partial\xi^2} + k_p^2 \right)\Phi = k_p^2\frac{a^2}{2}.
    \label{eq:one}
\end{equation}
Here $\xi$ is the spatial coordinate in the reference frame co-moving with the laser pulse; $k_p = \sqrt{ n_e e^2/m_e \varepsilon_0 c^2}$ is the plasma wave number with $n_e$ the ambient electron density, $e$ the elementary charge, $m_e$ the electron rest mass, $\varepsilon_0$ the vacuum permittivity, and $c$ the speed of light; $a =  eA/m_e c^2$ is the normalized vector potential of the driver. Equation~\eqref{eq:one} is valid for a drive pulse with a sufficiently wide spot size ($w_0 \gg 1/k_p$) and a very weak strength ($a\ll 1$) propagating in a low density plasma ($n_e \ll n_\mathrm{cr}$, $n_\mathrm{cr}$ denotes the critical density).

The resulting potential leads to a density perturbation of the form $n(\xi)\propto \sin\left[k_p(\xi-\xi_0)\right]$, where $\xi_0$ marks the center of the driver \cite{Esarey:2009ks}. Hence, the laser pulse sets up a sinusoidal density modulation in its wake with a period equal to the cold plasma wavelength 
\begin{equation}
    \lambda_p = 2\pi c \sqrt{\frac{m_e \varepsilon_0}{n_e e^2}}.
    \label{eq_plasma_wavelength_linear}
\end{equation}

With increasingly intense laser fields the wave excitation becomes non-linear and can be described by
\begin{equation}
    \frac{\partial^2 \Phi}{\partial\xi^2} = \frac{k_p^2}{2}\left[\frac{1+a^2}{(1+\Phi)^2}-1 \right].
    \label{eq_non_linear_waveequation}
\end{equation}

Analytical solutions for this equation exist only for specific laser profiles, e.g.~rectangular pulses\cite{Berezhiani:1990vl}. Importantly, the plasma wave train will have a different wavelength, $\lambda_{p,nl}$, in this non-linear regime. As summarized by \citet{Esarey:2009ks}, $\lambda_{p,nl}$ should scale according to
\begin{align}
    \label{eq_Esarey}
    \lambda_{p,nl}=\lambda_p 
    \begin{cases}
         1+3\chi^2/16 &\text{for } \chi\ll 1\\
         (2/\pi)(\chi+\chi^{-1}) &\text{for } \chi \gg 1
    \end{cases}
\end{align}
where the scaling parameter is $\chi=(a_0^2/2)/\sqrt{1+a_0^2/2}$. Thus, the plasma wavelength increases with intensity, which can be understood as a relativistic effect when electron oscillation in the plasma reaches relativistic energies. 

\begin{figure}[t]
    \begin{center}
    \includegraphics{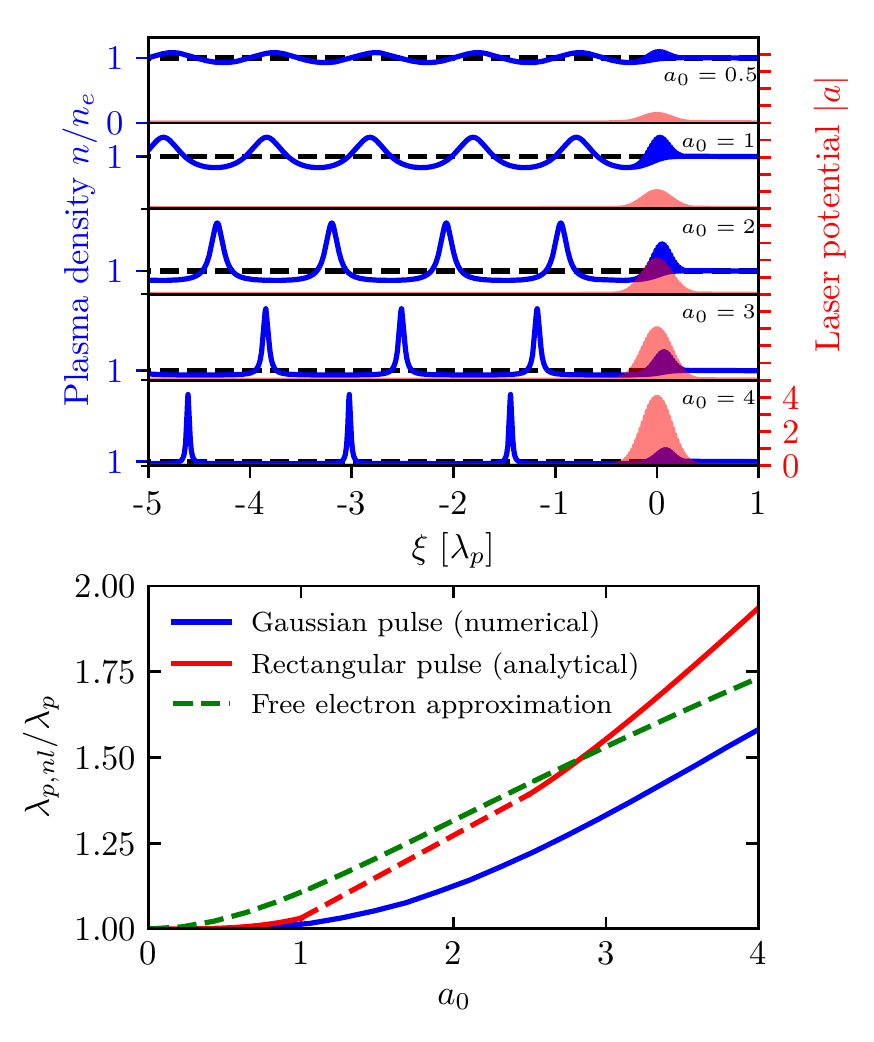}
    \end{center}
    \caption{\textit{Top:} Plasma waves excited by lasers with a gaussian envelope and different peak potential $a_0$ according to one-dimensional fluid theory, Eq.~\eqref{eq_non_linear_waveequation}. The transition from sinusoidal at low intensity ($a_0 < 1$) to increasingly non-linear density profiles ($a_0>1$) is clearly visible. \textit{Bottom:} Comparison of wavelength intensity dependence among different models. Note that the analytical expression for rectangular pulse, Eq.~\eqref{eq_Esarey}, has two disconnected region of validity. The dashed segment of the red curve is to guide the eye.}
    \label{fig:one}
\end{figure}

For more realistic pulse shapes, Eq.~\eqref{eq_non_linear_waveequation} needs to be solved numerically. In the upper panel of Fig.~\ref{fig:one}, such solutions are plotted for gaussian pulses with central wavelength $\lambda_0 = \SI{800}{\nano\meter}$, FWHM pulse duration $\tau=\SI{30}{\femto\second}$ and plasma density $n_e=\SI{3e18}{\per\cubic\centi\meter}$, corresponding to a ratio between pulse length and plasma wavelength of $c\tau/\lambda_p\approx0.5$. These parameters are chosen in accordance with our typical experimental conditions (see below), and again, the plasma wavelength shows a clear dependence on the intensity of the drive laser.

As the increase of the plasma wavelength for stronger lasers reflects the relativistic mass increase, \citet{Matsuoka:2010il} estimated $\lambda_{p,nl}$ from the momentum acquired by free electrons in the laser field. Based on the 1D assumption of a plane wave in which all electrons experience the peak potential $a_0$, the momentum acquired by the electrons is proportional to $a_0$, which leads to a modification of electron mass by the Lorentz factor $\gamma=\sqrt{1+a_0^2/2}$ in Eq.~\eqref{eq_plasma_wavelength_linear}. The factor of $1/2$ here reflects averaging over the fast oscillations assuming a linearly polarized driver. The non-linear plasma wavelength is therefore
\begin{equation}
    \lambda_{p,nl}=\lambda_p\left(1+a_0^2/2\right)^{1/4}.
    \label{eq_plasma_wavelength_Matsuoka}
\end{equation}

The lower panel of Fig.~\ref{fig:one} shows, as a function of the driver $a_0$, the variation of plasma wavelength predicted by models mentioned above. It can be seen that these models all predict an elongation of the plasma wave yet their values differ by more than 20\% for $a_0>2$.

The theoretical models make clear, though disparate, predictions for the intensity-dependent wavelength of the plasma oscillation. However, experimental confirmations have scarcely appeared in the literature. In previous experiments, \citet{Matlis:2006wg} reported curved wave fronts as evidence of a nonlinear plasma wave, yet due to the averaging effect of longitudinal probing geometry, no significant deviation from the linear wavelength was observed. Using a transverse geometry, \citet{Savert:2015jr} recorded the elongation of the plasma bubble, whereas the length of the second wave period remains consistent with the linear model until significant self-injection takes place, which suggests that beam-loading makes leading contribution to their observation of wave lengthening.

Contrary to the work by \citet{Savert:2015jr} where a single plasma bucket was studied, we take into account many oscillations, which significantly improves the measurement precision. Since injection is usually confined to the first few wakefield periods, averaging over many oscillations also suppresses the contribution of beam-loading hence allows us to measure the wavelength of free plasma oscillation with higher accuracy.

\section{Experiment}

\begin{figure*}[hptb]
\begin{center}
\includegraphics[width=.9\linewidth]{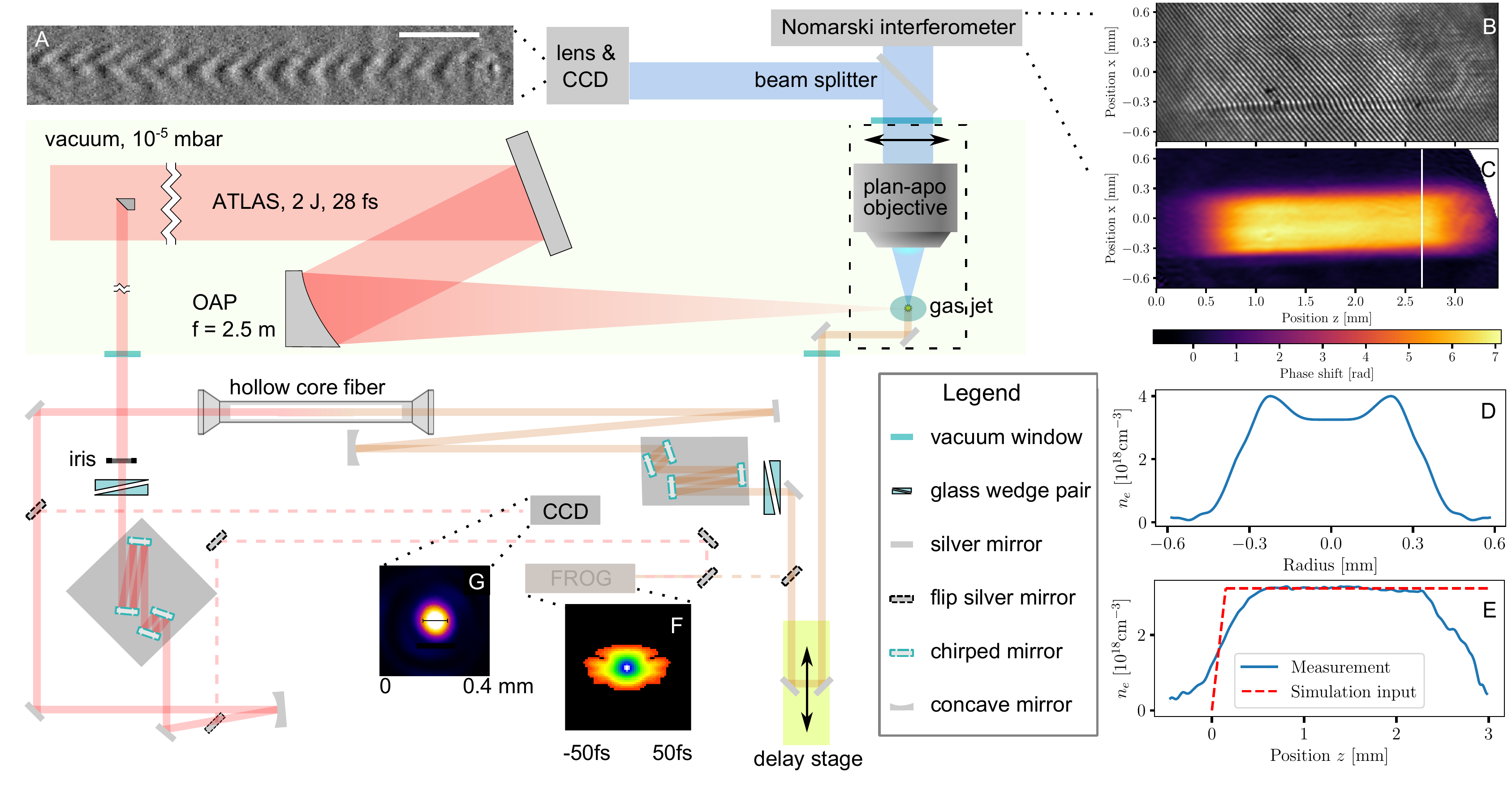}
\end{center}
\caption{Schematic representation of the experimental setup. \textit{Insets:}  \textbf{A)} An example few-cycle shadowgram of a nonlinear laser-driven plasma wave with a \SI{50}{\micro \meter} scale bar.
\textbf{B)} A raw image recorded with the Nomarski interferometer. \textbf{C)} Phase shift caused by the plasma, deduced from B. \textbf{D)} The transverse electron density profile retrieved from Abel-inversion at the position marked by the white line in C. Note that the density bumps at the shoulders and feet of the profile are a retrieval artifact. \textbf{E)} The longitudinal electron density profile at $x=0$ in C together with the density profile used for simulations (cf.~Figs.~\ref{fig:four}\&\ref{fig:five}). Note that the coordinate Position $z$ of the measured profile is shifted by \SI{-0.45}{mm} compared to C, and the plasma density of simulation input is scaled  to match the measurement. \textbf{F)} Retrieved FROG trace of the probe beam.  \textbf{G)} Far field profile of the probe beam measured with a CCD camera.}
\label{fig:two}
\end{figure*}

In the following we present a systematic comparison of the linear and nonlinear wakefields measured via optical probing. The experimental setup is schematically represented in Fig.~\ref{fig:two}. ATLAS is a Ti:sapphire laser system, delivering \SI{2}{\J} of pulse energy on target at \SI{5}{\Hz} repetition rate. With a central wavelength of $\lambda_0=\SI{800}{\nano\meter}$ and a FWHM bandwidth larger than \SI{50}{\nano \meter}, these pulses can be compressed to a FWHM duration of \SI{28}{\femto \second}, yielding a peak power of \SI{70}{\tera \watt}. A peak vacuum intensity of \SI{5.5e18}{\watt \per \square \centi \metre}, corresponding to $a_0 \simeq 1.6$, can be achieved at full power when an off-axis-paraboloid (OAP) with a focal length of \SI{2.5}{\metre} is used. Without the final amplifier, a peak power of \SI{13}{\tera \watt} is reached, which translates into $a_0 \simeq 0.7$ in vacuum. While these parameters are ideally suited to drive strong plasma waves\cite{Malka:2002eu}, the pulse duration of $\sim \SI{28}{ \femto \second}$ is too long for time-resolved probing in a perpendicular pump-probe geometry.

To generate the required few-fs probe pulses, a small fraction ($\sim$ \SI{1}{\milli \J}) of the ATLAS beam is picked off and coupled into an Ar-filled hollow core fiber. Self phase modulation (SPM) results in a spectrum spanning almost an octave, which, when compressed by an array of dispersive mirrors, leads to a transform limited pulse duration of less than \SI{10}{\femto \second} (see appendix for more details on the setup). The few-cycle pulses are then sent transversely through the interaction region and collected by a plan-apochromatic microscope objective to form either shadowgrams or interferograms with a spatial resolution of $\sim$ \SI{2}{\micro \metre}. The probe beam including the imaging setup can be moved with respect to the gas target along the main laser axis without changing the relative delay, which allows different parts of the target to be sampled.

\begin{figure*}[bth]
\begin{center}
\includegraphics[width=.9\linewidth]{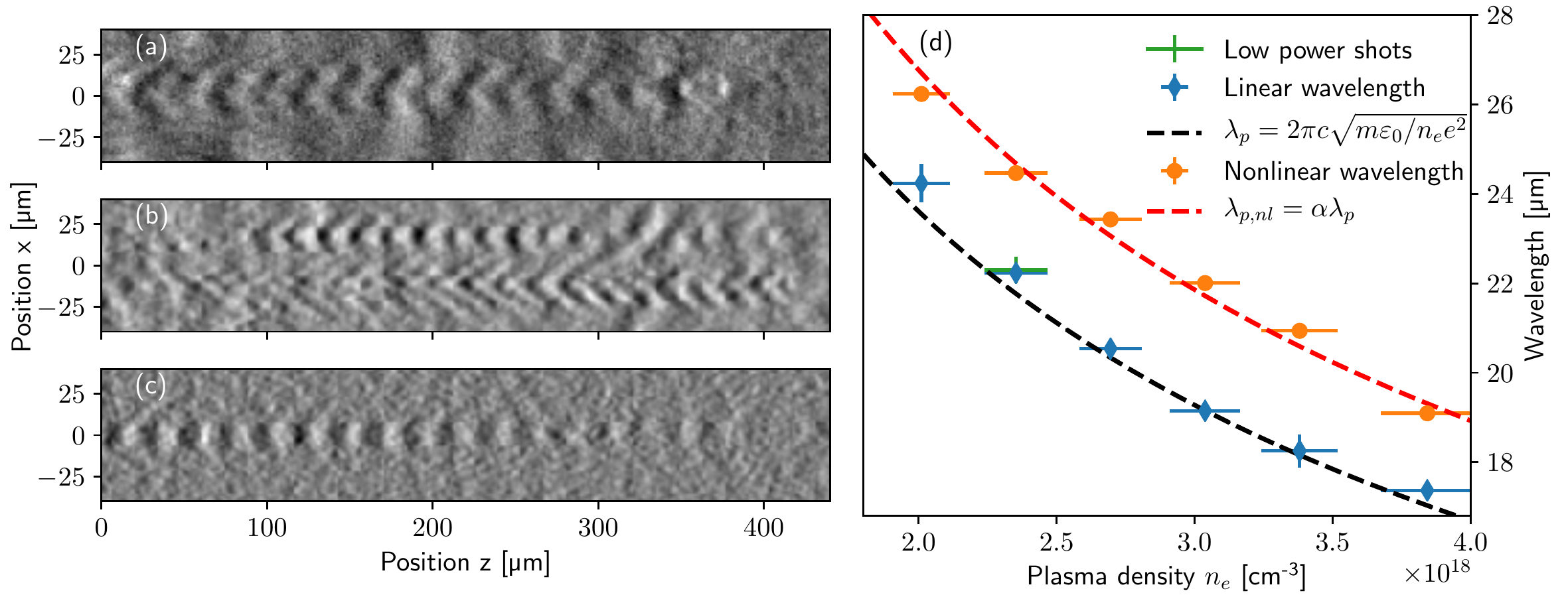}
\end{center}
\caption{\textit{Left:} Representative shadowgrams of laser driven plasma waves in the plasma density range of $n_e=2-\SI{4e18}{\per\cubic\centi\meter}$. (a) a nonlinear plasma wave driven by a \SI{70}{\tera\watt} pulse. (b) a strongly nonlinear plasma wave driven by a \SI{70}{\tera\watt} pulse with a weaker secondary wave above it. Note that the secondary wave starts at the same position as the main wave, but its modulation at the front is poorly visible due to its overlap with a diffraction feature in the probe's near-field profile. (c) a quasi-linear plasma wave driven by a \SI{13}{\tera\watt} pulse. \textit{Right:} The wavelength of plasma oscillation as a function of electron density. The nonlinear wavelengths (orange dots) are obtained from the main waves whereas the linear wavelengths (blue dots) are deduced from the filaments, cf.~panel (b). The low power shots (green dot) are taken at \SI{13}{\tera\watt}, cf.~panel (c). Each data point is an average of 2 to 9 shots. The vertical error bars represent the standard error of mean (s.e.m.) of each run. The horizontal error bars are the estimated uncertainties in the density retrieval from interferometry. A least square fit to the nonlinear wavelengths (dashed red line) yields the elongation factor $\lambda_{p, nl}/\lambda_p$ of $\alpha = 1.13$.}
\label{fig:three}
\end{figure*}

Figure~\ref{fig:three}~(a) shows a shadowgraphic snapshot of a plasma wave driven by a laser pulse at full pulse energy ($\SI{70}{\tera\watt}$ on target, vacuum $a_0\simeq 1.6$). The variation of local plasma density in the wakefield imprints a position-dependent phase on the probe beam, which leads to a modulation of the probe beam intensity after propagation. Since the intensity modulation is proportional to the second derivative of the phase distribution, the larger-scale features of the plasma can be identified. First and foremost, the periodicity of the modulation reflects the local plasma wavelength and the wave fronts are curved, implying a nonlinear wave. 

Occasionally, we observe secondary plasma waves with shorter wavelengths in the vicinity of the main wave, cf.~Fig.~\ref{fig:three}~(b), which we interpret as signs of filamentation. This is likely due to the laser being slightly out of focus at the gas jet edge and its mid/far-field intensity distribution being imperfect. As those filaments are expected to have lower intensities than the main focus, this observation hints at an intensity-dependent plasma wavelength. We therefore carried out measurements at the same density but at reduced laser power ($\SI{13}{\tera\watt}$ on target, vaccum $a_0\simeq 0.7$) and measured plasma waves with significantly shorter wavelength. In fact the plasma wavelength of low power shots was similar to that of the filaments, cf.~Fig.~\ref{fig:three} (c).

To establish a quantitative relation between the measured plasma wavelengths and the non-relativistic model Eq.~\eqref{eq_plasma_wavelength_linear}, we independently determined the electron density by Normaski-type interferometry. Owing to the large field of view of the interferometry camera, this provided an in-situ measurement of the phase difference between the plasma column ionized by the drive laser and the background gas in the jet.
The density can then be retrieved via Abel inversion, assuming a cylindrical symmetry of the plasma channel. 

We performed such density measurements in the center of a 3-mm-long hydrogen gas jet (about \SI{1.8}{mm} of propagation in plasma) and simultaneously recorded the shadowgrams of laser-driven plasma waves. Within a density range of $n_e=2-\SI{4e18}{\per\cubic\centi\meter}$, the wavelength deduced from shadowgrams is $\lambda_{p,nl}=1.13\lambda_p$ for the main wave at full power. In contrast, the wavelength of both the filament- and low-power driven waves does not significantly differ from the expected cold $\lambda_p$, cf.~Fig.~\ref{fig:three}~(d).

It should be noted that the intensity modulation in shadowgrams taken at plasma densities of $n_e\sim10^{18}\:\si{\per\cubic\centi\meter}$ is not entirely of the same nature compared with that at higher densities e.g.~$n_e\sim10^{19}\:\si{\per\cubic\centi\meter}$ as reported by \citet{Savert:2015jr}. Thanks to the high gradient of the refractive index, \citet{Savert:2015jr} could image the plasma-wave-induced intensity modulation of the probe beam in the plane of the drive laser, whereas our measurements showed only weak contrast at this position. Instead, at a distance of $\sim\SI{100}{\micro\meter}$ away from laser axis, we could observe stronger intensity modulation. That is to say, our shadowgraphy technique is essentially propagation-based phase-contrast imaging. As such, the intensity contrast of a single plasma bucket is given by the distance of the wakefield and the image plane and therefore susceptible to the shot-to-shot pointing fluctuation and the long-term drift of the drive laser in our experiment. Consequently, we cannot reliably measure the bubble size, but can only retrieve the wavelength of periodic features.

\section{Discussion}
According to the models from Section I, it should be straightforward to deduce the local peak potential $a_0$ from the measured plasma wave elongation. In the \SI{13}{\tera\watt} case, all models predict an elongation of $1-2\%$, which is within the measurement uncertainty. However, it turns out that relating the measured $\lambda_{p,nl}$ to a realistic value of driver $a_0$ is much more difficult at \SI{70}{\tera\watt}, as is summarized in Table~\ref{tab:comparison}. 

\begin{table}[hbt]
    \centering
\begin{tabular}{l|c} \hline
Method     &  Estimated $a_0$\\ \hline
1D non-linear model for rectangular pulse     & 1.6 \\
1D non-linear model for gaussian pulse    & 1.95 \\
Momentum based estimate & 1.15 \\
Momentum based estimate (FWHM average) & 2.15 \\ \hline
Vacuum focus & 1.6 \\
Matched spot size  & 4.0 \\
Particle-in-cell simulation & 4.5\\
\hline
\end{tabular}
\caption{\textit{Upper part:} Various estimates for the laser $a_0$ based on models for the non-linear plasma wavelength (cf.~Sec.~I) and the measured value $\lambda_{p,nl}=1.13\lambda_p$. \textit{Lower part:} Comparison with estimates based on the measured focal spot and pulse energy, the matched spot size for $P=\SI{70}{\tera\watt}$ and $n_e=\SI{3e18}{\per\cubic\centi\meter}$ and the result from a PIC simulation after 2 mm of propagation.}
    \label{tab:comparison}
\end{table}

\begin{figure*}[ht]
\begin{center}
\includegraphics[width=.95\linewidth]{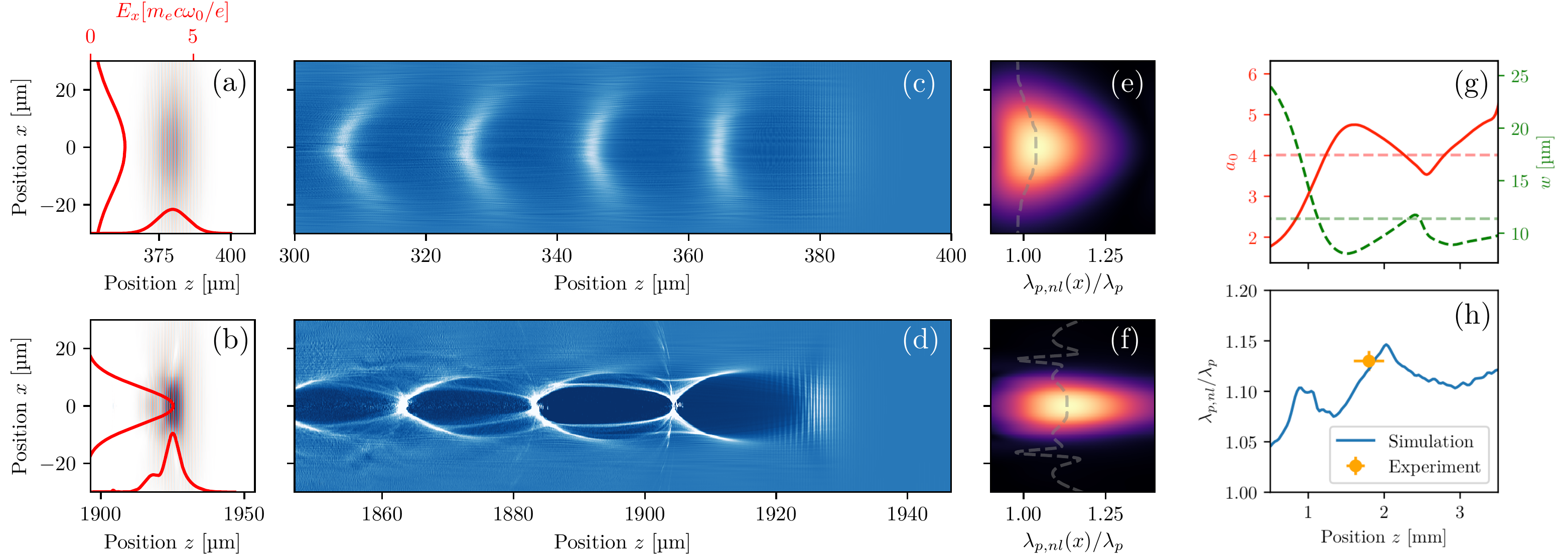}
\end{center}
\caption{Snapshots of a quasi-3D simulation of a \SI{70}{\tera \watt}\, \SI{30}{fs} (FWHM) pulse propagating in a 3-mm-long hydrogen gas jet with a nominal electron density of $n_e=\SI{3e18}{\per\cubic\centi\meter}$: upper panels are for the beginning of the jet and lower panels are for the center of the jet where the experimental data are taken (cf.~Fig.~\ref{fig:three}). From left to right: (a-b) the intensity distribution (false color) together with the E-field envelope of the laser pulse in transverse and longitudinal direction (red lines), normalized by $m_e c \omega_0/e$, with $\omega_0$ the laser carrier frequency. (c-d) the electron density distribution. (e-f) line-by-line Fourier transform of the electron density with the abscissa converted from wave number to wavelength and the intensity corrected by the Jacobian (false color), and the position of the intensity maximum at each transverse coordinate $x$ (the dashed line). Note that the wiggles in (f) are a numerical artifact due to the weak density modulation outside the drive laser. (g) the evolution of the peak laser potential (red solid line) and the beam waist (green dashed line). The horizontal lines indicate the matched condition from \citet{Lu:2006ja}. (h) the evolution of the elongation factor (blue line), which shows good agreement with the measurement (orange dot). The vertical error bar of the measured dot indicates the 95\% confidence interval of the elongation estimate and the horizontal error bar is the sum in quadrature of the length of the visible wave train and the uncertainty in determining the length of the gas jet up-ramp.}
\label{fig:four}
\end{figure*}

\begin{figure*}[ht]
\begin{center}
\includegraphics[width=.9\linewidth]{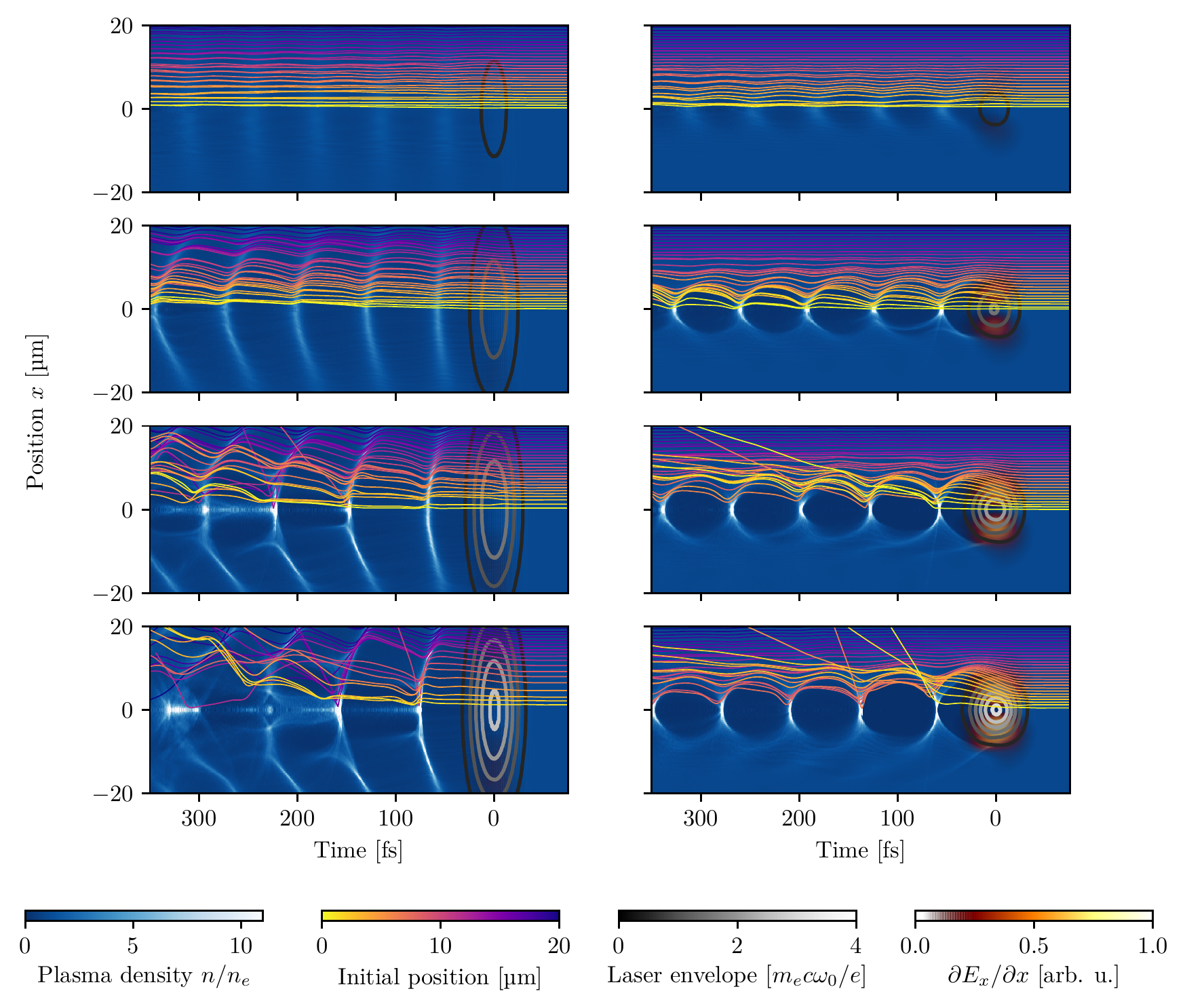}
\end{center}
\caption{Comparison of plasma wave train formation in the wake of a tightly focused spot (right) and a 3 times as wide spot (left) at various laser peak potentials ($a_0=[1.0,2.0,3.0,4.0]$ from top to bottom). Colored lines show the trajectories of electrons with different initial position in radial coordinates. The normalized E-field strength of the laser is shown as grey-scale contour plot and the transverse field gradient is indicated with an overlaid colormap. The transverse gradient of the tightly focused laser leads to stronger transverse electron motion and thus prevents them from experiencing the peak laser potential. This further leads to cavitation and suppresses the elongation of the wave train. In contrast, the simulations for a wide focal spot are comparable to laminar models, with a characteristic horseshoe-like shape, until wavebreaking sets in for $a_0\gtrsim 3$. All simulations are performed for a plasma density $n_e = \SI{3e18}{\per\cubic\centi\meter}$ and using an FWHM pulse duration of $\tau = \SI{30}{\femto\second}$.}
\label{fig:five}
\end{figure*}

We first tried to interpret the measured elongation factor of 1.13 using the analytical solution of 1D fluid theory with a square pulse, cf.~Eq.~\eqref{eq_Esarey}, which yields a laser peak potential $a_0=1.6$ and a scaling parameter $\chi=0.85$. Note that the scaling factor $\chi$ considerably differs from the model's validity range ($\chi \ll 1$), rendering this result rather unreliable. Given the experimental pulse shape cannot be considered a rectangle anyway, we next used a more realistic gaussian pulse. The numerical solution of Eq.~\eqref{eq_non_linear_waveequation} suggests a slightly higher value of $a_0=1.95$.

On the other hand, we can associate the observed lengthening with the relativistic increase of the electron mass, yielding a $\gamma$-factor of 1.28 at full power. Based on Eq.~\eqref{eq_plasma_wavelength_Matsuoka} this would correspond to a normalized potential of 1.12, even lower than the expected vacuum potential. However, the assumption that all electrons experience the same intensity, i.e.~the peak potential $a_0$, is unrealistic. Instead, the retrieved value should be interpreted as an averaged potential $\langle a \rangle$. If we assume a gaussian shape of the intensity profile and take the average within the full width at half maximum in both transverse and longitudinal directions, a peak value of $a_0=2.15$ is obtained. 

To sum up, there is a large variation between the estimates from the models discussed in Section I. The results are roughly compatible with the vacuum focus intensity, but the laser will self-focus inside the plasma and we therefore expect a much higher value for $a_0$ inside the plasma. For a $\SI{70}{\tera\watt}$ laser in a plasma with $n_e=\SI{3e18}{\per\cubic\centi\meter}$ we estimate a matched spot size $w_0=\SI{12}{\micro\meter}$ and a peak potential $a_0=4.0$ from the model by \citet{Lu:2006ja}, if we neglect the energy deposition to the plasma.
Hence the $a_0$ values deduced from the measured elongation factor using the models plotted in Fig.\ref{fig:one} are significantly too small.
On the other hand, as no external guiding technique is applied and the initial spot size does not fulfill the self-guiding condition, the laser is also expected to evolve strongly during the propagation. Consequently, there is considerable uncertainty in the driver intensity at the measurement point. 

In order to understand the experimental results in detail, and to gain insight on the evolution of the drive laser,
we have used the quasi-3D code FBPIC \cite{Lehe:2016dn} to simulate the laser propagation and plasma wave formation. Similar to other quasi-3D codes such as CALDER-CIRC\cite{Lifschitz20091803}, FBPIC employs an azimuthal Fourier decomposition, where the lowest two modes are associated with the radial symmetric component of the wakefield and the laser field, respectively. As the wake can become asymmetric at large laser intensities, higher order modes $m>2$ might become necessary to model the system\cite{Lifschitz20091803}. Here we used $m=4$ modes, with a resolution of $\Delta z = \lambda_0/30$ in longitudinal and $\Delta r \approx \lambda_p/100$ in the radial direction for a simulation window of $z\times r = (100\times 65)\:\si{\square\micro\meter}$, initialized with 32 particles per cell for $r<\SI{30}{\micro\meter}$. The plasma is considered as completely pre-ionized with a longitudinal density profile consisting of a $\SI{150}{\micro\meter}$ linear ramp, followed by a constant density of $n_e=\SI{3e18}{\per\cubic\centi\meter}$. For the driver, we set up a laser pulse in vacuum with a FWHM duration $\tau = \SI{30}{\femto\second}$, a FWHM spot size of $\SI{30}{\micro\meter}$, and a peak potential $a_0=1.6$. The simulation results are summarized in Fig.~\ref{fig:four}.

At the beginning of the gas jet, the \SI{70}{TW} laser pulse has a FWHM spot diameter of \SI{30}{\micro \metre}, larger than its FWHM pulse length $c\tau = \SI{9}{\micro \meter}$ and the linear plasma wavelength $\lambda_p = \SI{19.3}{\micro\meter}$, hence the plasma motion is still predominantly \textit{longitudinal}. As a result, the lengthening of the plasma wave train follows roughly the laser's radial intensity distribution and the wave fronts become curved with the curvature increasing farther behind the driver, cf.~Fig.~\ref{fig:four} (e) and (c). 

Over the first millimeters of propagation, self-focusing reduces the spot size to below $\lambda_p$ and the transverse component of the ponderomotive force becomes comparable to its longitudinal one. In this case, transverse plasma oscillations cause complete electron cavitation behind the driver, leading to the well-known bubble-like structure\cite{Pukhov:2002ts}. Furthermore, comparing the dashed lines in Figs.~\ref{fig:four} (e,f), the plasma wavelength ceases to vary in the transverse direction and appears to be almost a constant throughout the wave due to phase mixing of plasma oscillation.  

Seen from the rightmost column of Fig.~\ref{fig:four},
the peak laser intensity oscillates in the range $a_0 = 3.5$ to $4.5$, which encompasses the matched value of $a_0 = 4$; this oscillation is caused by the unmatched initial spot size. At the same time, the lengthening of the plasma wave train $\lambda_{p,nl}/\lambda_p$ is between $1.10-1.15$, much smaller than the prediction of the models in Fig.~\ref{fig:one} for a pulse with $a_0 \sim 4$, yet compatible with the experimental data. Indeed, at the center of the jet, the simulation accurately reproduces the measured plasma wave lengthening of 13\%.

The poor performance of the widely established analytical models compared to the good agreement between simulation and experiment therefore indicates that the physics of plasma wave trains is governed by effects that are not included in the model systems. In particular, it is too simplistic to assume that the wave train formation is dominated by a single parameter, the peak potential $a_0$.
As the plasma wave is generated by the ponderomotive force\cite{Esarey:2009ks}, $\vec F_p=-m_ec^2\vec \nabla(a^2/2)$ (for $a_0\ll1$), which depends on the gradient of the intensity, the wave formation will not only depend on the peak value $a_0$, but also the pulse length and width. The latter is particularly important, as it directly influences the \textit{transverse} motion of electrons and therefore plays a major role in the breakdown of any laminar, one-dimensional model.

To illustrate this behavior, Fig.~\ref{fig:five} shows simulation results for both the plasma waves and the trajectories of plasma electrons driven by laser pulses of different peak intensities and widths. We compare a pulse with a spherical intensity contour $w_0=c\tau$ on the right to a laser with a larger focus, i.e.~an aspect ratio $w_0/c\tau=3$, as shown in the left column.

As expected, in the case of a wide focal spot, a simple extension of the 1D theory to higher dimensions assuming laminar motion still seems reasonable for peak potentials $a_0<3$. The wave amplitude and wavelength are modulated by the radial intensity profile of the laser, leading to a horse-shoe like structure in the laser's wake. With peak potentials $a_0 >3$, the fields reach the (relativistic) wave-breaking limit and hence the fluid model breaks down, marked by the self-injection into the wakefield.

For the tightly focused case, the fluid model breaks down even sooner and the dynamics of the plasma wave fundamentally change. Due to the increasingly transverse motion, many electrons drift farther away from laser axis even before they experience the local intensity peak of the laser pulse. As a result, a high proportion of the plasma electrons involved in the wake formation do not experience the peak laser potential, reducing their oscillation strength.
The fluid model therefore fails in this case for $a_0 \gtrsim 1$.

\begin{figure}[ht]
\begin{center}
\includegraphics[width=.9\linewidth]{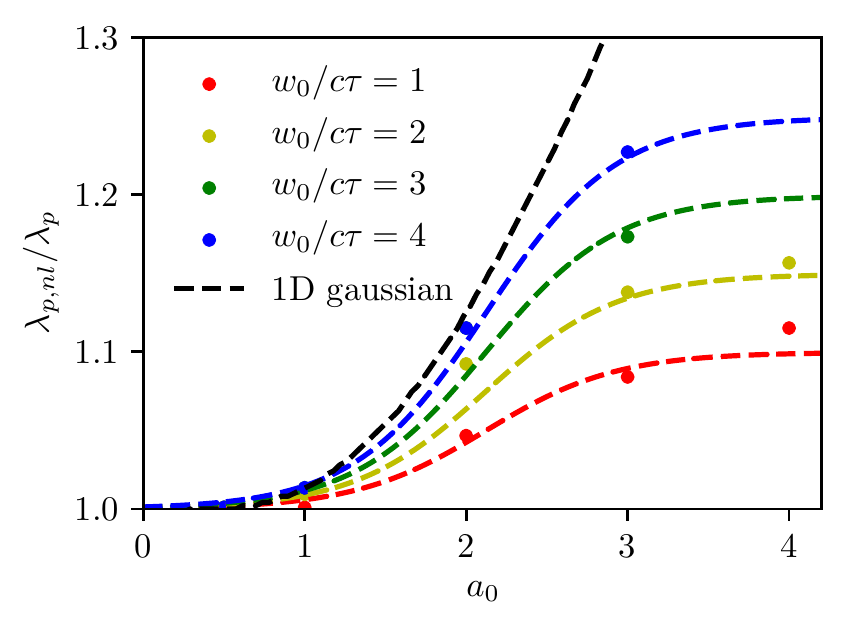}
\end{center}
\caption{Plasma wave elongation according to PIC simulations with different aspect ratios $w_0/c\tau$ of the laser pulse (dots). The fit function (colored dashed lines) agrees well with the simulations, while approaching the 1D non-linear model for $w_0/c\tau\gg1$ and $a_0\lesssim 2$. }
\label{fig:six}
\end{figure}

To quantify these effects, we have performed a total of 20 simulations with $a_0=0.5-4.0$ and varying aspect ratios $w_0/c\tau=1-4$ of the laser pulse, cf.~Fig.~\ref{fig:six}. 

Within our parameter range, $\Delta\lambda=\lambda_{p,nl}-{\lambda_p}$ obtained from simulation data can be reasonably well described with a sigmoid function along $a_0$, while the wavelength also increases proportionally to the aspect ratio $w_0/c\tau$:
\begin{equation}
\Delta\lambda(a_0,w_0)=\frac{p_0}{1 + e^{-  p_1\cdot(a_0-p_2)}}\times\left(1+\frac{w_0}{c\tau}\right).
\end{equation}
A least-squares fit yields the parameters $p_0=0.05$, $p_1=-2.5$, and $p_2=2.1$. The sigmoid's midpoint $p_2$ of this fit function can be taken to be the value of $a_0$ at which damping becomes significant.
This damping, which is absent in the one-dimensional case, can be explained by the aforementioned effect that the plasma wave is mainly formed by electrons from outer radii. These electrons only interact with the outer part of the laser where the potential is $a\sim1-2$. An increase in $a_0$ only moves these zones further outwards and leads to a slightly larger (transverse) bubble size, but it does not substantially increase the peak intensity experienced by the plasma electrons which form the wake. In contrast, for a wider laser the average field experienced by plasma electrons is higher and thus, the plasma wavelength will increase with the aspect ratio.

For even wider drivers, the electron motion becomes more and more longitudinal and the plasma wave formation gradually approaches the solution to the 1D non-linear wave equation~\eqref{eq_non_linear_waveequation}, and the elongation factor can be  $\lambda_{p,nl}/\lambda_p>1.2$. 
On the contrary, the plasma waves driven by tightly (self-)focused drivers clearly differ from the models plotted in Fig.~\ref{fig:one}, and show only a weak increase of the order of 10\% in the plasma wavelength, as $a_0$ is increased. It is important to note that this behavior is not covered by the model of \citet{Lu:2006ja}, which only concerns the \textit{first} bubble and not multi bubble trains as shown in Fig.~\ref{fig:five}\footnote{The main problem here is that the electrons which form the first bubble are not that same which oscillate in the second and third.}. It is therefore important to develop new theoretical models for this regime of wakefield formation, which is central to many experiments such as multi-pulse wakefield excitation\cite{Hooker:2014ij,Cowley:2017dz}.

\section{Conclusion}
In conclusion, we have presented in-situ measurements of linear and non-linear plasma waves driven by a 100-TW-class laser system in the typical plasma density range for GeV-class laser wakefield accelerators. The combination of shadowgraphic snapshots of the plasma waves with interferometry allowed us to measure 
elongation of the plasma wavelength by up to 13\%.
% a 13 percent elongation in the plasma wave train period. 
% This is contrary to predictions of established analytical models, yet, reproduced in quasi-3D PIC simulations.
These experimental results were found to be inconsistent with analytical models, but in agreement with quasi-3D PIC simulations.

Our analysis shows that in addition to the peak laser potential $a_0$, the spot size of the laser is a dominant scaling parameter for the plasma wavelength $\lambda_{p,nl}$ due to its effect on the transverse ponderomotive force. As a consequence the plasma wavelength cannot be used as a direct diagnostic for the laser intensity without knowledge about the focus size at the probed position. Future studies aiming for in-situ measurements of the laser intensity will therefore need to develop additional diagnostics to measure the self-focused size of the laser. Alternatively, one can implement a non-invasive beam diagnostic by means of short, low-density gas jets with negligible self focusing. The method can also be applied to study future wakefield accelerator schemes, such as multi-pulse wakefield accelerators or beam-driven wakefield accelerators\cite{Hidding:2019fa}.

While an empirical model has been used to fit simulation results in our parameter regime, it remains to be explored how the non-linear wave formation changes as function of other parameters such as the background plasma density and laser pulse duration. Such multi-parameter studies cannot be based on particle-in-cell simulations due to computational costs and hence, further studies will require reduced numerical or enhanced analytical models. In particular, we have identified several effects that contribute to the observed scalings of the plasma wavelength, e.g.~the effective potential experienced by electrons forming the plasma wave and the transition between longitudinal and transverse plasma oscillations.

We would like to emphasize that most laser wakefield accelerators operate neither in the quasi-linear, laminar regime, nor in the transverse bubble regime, but rather at the transitional regime explored in this work. We hope that our results will motivate further analytical studies to understand the transitional regime of laser wakefield formation. Last, it should also be noted that our analysis ignores beam-wakefield interaction. Further research will be required to take the effects of beamloading into account, which also have an impact on the wakefield formation.

\section*{Appendix}

\textbf{Few-cycle pulse generation.} An half-inch mirror clips out about \SI{10}{\milli \J} of the ATLAS beam. It is then guided through a \SI{1}{mm} thick fused silica window to a probe table outside the vacuum  target chamber (cf.~Fig.~\ref{fig:two}). An iris and ND filters adjust the diameter and the energy of the probe pulse to about \SI{8}{mm} and \SI{1}{\milli \J}, respectively. A dispersive mirror array together with a variable thickness glass wedge pair compensates the group delay dispersion (GDD) accumulated during pre-fiber propagation and therefore ensures effective self phase modulation (SPM) inside the Ar-filled hollow core fiber. The installed dispersive mirrors provide a nominal GDD  of \SI{-40}{\square \femto \second} per reflection for the p-polarized light in the spectral range of \num{500} - \num{1050} \si{\nano \metre} and need to be used in pairs with incidence angles of 5 and 19 degrees. The hollow core fiber in this setup has an inner-diameter of \SI{240}{\micro \metre} and a length of \SI{0.9}{\metre}. With a filling pressure of \SI{500}{\milli \bar}, about \SI{400}{\milli \J} can be transmitted though the fiber. A second array of dispersive mirrors and a wedge pair compress the pulse then close to its Fourier limit. A motorized delay stage ensures proper synchronization and allows to study the plasma wave evolution by setting the relative delay between main pulse and probe pulse.

\textbf{Shadowgraphy and image treatment.} After passing through the plasma region, the probe beam is imaged directly onto a CCD camera using an infinity conjugate NIR plan-apochromatic microscope objective together with an achromatic lens. Flat-field correction are performed for the images using timely acquired background. Low-order regressions are employed in longitudinal and transverse directions to remove inhomogeneities of the probe beam intensity. After this initial image treatment, FFT is then calculated for each slice along the direction of laser propagation in the region of interest. To obtain the wavelength from the frequency, the Jacobian conversion is used.

\textbf{Interferometry measurements.} To independently deduce the plasma density in the experiment, a Nomarski type interferometer is used, see also Ref.\cite{Guillaume:2015dia}. More specifically, a Wollaston prism with 1$^{\circ}$ separation angle is installed in one arm of the probe after the beam splitter, followed by a polarizer, a lens to reduce the magnification and an interferene filter transmitting \SI{880 \pm 5}{\nano \metre}. A second polarizer in front of a CCD camera enbales the interference. Based on the numerical aperture of the last lens, a resolution of about \SI{10}{\micro \metre} is estimated. Phase retrieval is performed using the IDEA software kit \cite{Hipp:2004IDEA} and plasma densities are estimated using Abel inversion with the Backus-Gilbert method. 

\textbf{Temporal evolution of the plasma wavelength.} It is worth noting that the presented method to directly compare the wavelength of the wave train with the plasma density is only valid for a constant plasma density, e.g.~along the density plateau of a supersonic gas jet. In presence of density gradients the phase slippage over time between the adjacent plasma oscillations of different frequency has to be taken into account\cite{Brantov:2008dm}, i.e.~
\begin{equation}
    {\lambda_p(z,t)} = {\lambda_p(z)}{\left(1-\frac{(z-ct)}{\lambda_p(z)}\frac{d\lambda_p(z)}{dz}\right)^{-1}}.
\end{equation}

\textbf{Particle-in-cell simulations.} For quasi-3D simulations we use the PIC code FBPIC\cite{Lehe:2016dn}, which uses a spectral cylindrical representation. All simulations are performed at a density of $n_e=\SI{3e18}{\per\cubic\centi\meter}$. For the parameter scan we use $m=2$ modes and a resolution of $\Delta z = \SI{40}{nm}=\lambda_0/20$ and $\Delta r = \SI{200}{nm}\approx \lambda_p/100$ in longitudinal and radial direction, respectively. The simulation box has a length of $\SI{150}{\micro\meter}$ and each simulation is stopped after 0.2 mm of propagation in order to avoid laser evolution effects such as self-focusing or self-compression.

\section*{Acknowledgments}
This work was supported by DFG through the Cluster of Excellence Munich-Centre for Advanced Photonics (MAP EXC 158), TR-18 funding schemes, by EURATOM-IPP, the Center for Advanced Studies of the Ludwig-Maximilians-Universit{\"a}t M{\"u}̈nchen and the Max Planck Society. The authors are grateful to the Gauss Centre for Supercomputing e.V. (www.gausscentre.eu) for funding this project by providing computing time on the GCS Supercomputer SuperMUC at Leibniz Supercomputing Centre (www.lrz.de).

\end{document}